\documentclass[10pt,letterpaper]{article}
\usepackage[top=0.85in,left=2.75in,footskip=0.75in]{geometry}

\usepackage{amsmath,amssymb}

\usepackage{changepage}

\usepackage[utf8x]{inputenc}

\usepackage{textcomp,marvosym}

\usepackage{cite}

\usepackage{nameref,hyperref}

\usepackage[right]{lineno}

\usepackage{microtype}
\DisableLigatures[f]{encoding = *, family = * }

\usepackage[table]{xcolor}

\usepackage{array}

\newcolumntype{+}{!{\vrule width 2pt}}

\newlength\savedwidth

\newcommand\thickhline{\noalign{\global\savedwidth\arrayrulewidth\global\arrayrulewidth 2pt}%
\hline
\noalign{\global\arrayrulewidth\savedwidth}}

\raggedright
\usepackage[aboveskip=1pt,labelfont=bf,labelsep=period,justification=raggedright,singlelinecheck=off]{caption}

\makeatletter
\renewcommand{\@biblabel}[1]{\quad#1.}
\makeatother

\usepackage{lastpage,fancyhdr,graphicx}
\usepackage{epstopdf}
\fancyheadoffset[L]{2.25in}
\fancyfootoffset[L]{2.25in}
\usepackage{bm}  
\usepackage{algpseudocode,algorithmicx}
\newcommand*\Let[2]{\State #1 $\gets$ #2}
\algrenewcommand\algorithmicrequire{\textbf{Precondition:}}
\algrenewcommand\algorithmicensure{\textbf{Postcondition:}}
\newcommand{\boldtheta} {\boldsymbol{\theta}} 
\newcommand{\abs}[1]{\vert#1\vert} 
\newcommand{\sign}[1]{\mathrm{sign}(#1)}
\newcommand{\sd}[1]{\mathrm{sd}\left(#1\right)}

\newcommand{\firstuse}[1]{\emph{#1}} 
\newcommand{\ie}{i.e.\ }
\newcommand{\eg}{e.g.\ }
\newcommand{\latin}[1]{\textit{#1}}
\newcommand{\etal}{et al.\ }

\newcommand{\ignore}[1]{}
\newcommand{\unix}[1]{\texttt{#1}} 

\begin{document}
\vspace*{0.2in}

\begin{flushleft}
{\Large
\textbf\newline{Exponential random graph model parameter estimation for very large directed networks} 
}
\newline
\\
Alex Stivala\textsuperscript{1,2*},
Garry Robins\textsuperscript{3},
Alessandro Lomi\textsuperscript{1,4}
\\
\bigskip
\textbf{1} Institute of Computational Science, Università della Svizzera italiana, Lugano, Ticino, Switzerland
\\
\textbf{2} Centre for Transformative Innovation, Swinburne University of Technology, Hawthorn, Victoria, Australia
\\
\textbf{3} Melbourne School of Psychological Sciences, The University of Melbourne, Victoria, Australia
\\
\textbf{4} The University of Exeter Business School, Rennes Drive, Exeter, United Kingdom
\\
\bigskip

%
%





* alexander.stivala@usi.ch

\end{flushleft}

\section*{Abstract}
Exponential random graph models (ERGMs) are widely used for modeling
social networks observed at one point in time. However the
computational difficulty of ERGM parameter estimation has limited the
practical application of this class of models to relatively small
networks, up to a few thousand nodes at most, with usually only a few
hundred nodes or fewer. In the case of undirected networks, snowball
sampling can be used to find ERGM parameter estimates of larger
networks via network samples, and recently published improvements in
ERGM network distribution sampling and ERGM estimation algorithms have
allowed ERGM parameter estimates of undirected networks with over one
hundred thousand nodes to be made.  However the implementations of
these algorithms to date have been limited in their scalability, and
also restricted to undirected networks.  Here we describe an
implementation of the recently published Equilibrium Expectation (EE)
algorithm for ERGM parameter estimation of large directed networks. We
test it on some simulated networks, and demonstrate its application to
an online social network with over 1.6 million nodes.



\section*{Introduction}

Exponential random graph models (ERGMs) are a class of statistical
model often used for modeling social networks
\cite{lusher13,amati18}. Parameter estimation in these models is a
computationally difficult problem, and algorithms based on Markov
chain Monte Carlo (MCMC) are generally used
\cite{coranderetal98,coranderetal02,snijders02,hunter06,robins07,caimo11,hummel12,hunter12,amati18}. The
computational time required by these methods places a limit on the
size of networks for which models can be estimated in practice. A
recently published new algorithm for sampling from the ERGM
distribution can reduce this time by an order of magnitude
\cite{byshkin16}, and a new estimation algorithm even more
\cite{byshkin18}, however scalability is still
a problem when extremely large networks are considered. It is also worth
noting that the state space for a directed network is far larger than
for an undirected network with the same number of nodes
\cite{robins09}, and so this problem is even more difficult in the
case of directed networks.

One solution to this problem is to take snowball samples
\cite{coleman58,goodman61,goodman11,heckathorn11,handcock11} from the
original network, and estimate ERGM parameters from these
\cite{handcock10,stivala16}. The first description of such a method
was \cite{handcock10}. However this method requires that estimation
over the entire set of random tie variables is feasible, limiting the
size of networks to which the method can be applied in practice.
A more recently proposed method \cite{stivala16} is to estimate
parameters for each sample in parallel with conditional estimation
\cite{pattison13}, combining the estimates with a meta-analysis
\cite{snijders03} or using bootstrap methods \cite{efron87} to
estimate the standard errors. This work, however, was only applied to
undirected networks. 

However the problem of directly estimating ERGM parameters for a very
large network (rather than from snowball samples) remains,
particularly for directed networks where snowball sampling is not
straightforward. Here we describe an implementation of the
Equilibrium Expectation (EE) method \cite{byshkin18} extended to
directed networks, which is scalable and efficient enough to be used
to estimate ERGM parameters for networks with over one million nodes.

Hunter \& Handcock \cite[p.~581]{hunter06} note that the largest
network estimated to date (in 2006) was the $N=2209$ nodes adolescent
friendship network estimated by Hunter, Goodreau, \& Handcock
\cite{hunter08}. However this network was treated as undirected. Larger
undirected networks subsequently had ERGM models estimated indirectly
by snowball sampling, with the largest having 40~421 nodes
\cite{stivala16}. By using an improved ERGM distribution sampler,
Byshkin \etal \cite{byshkin16} could directly estimate ERGM parameters
for a 3061 node patient transfer network (treated as undirected),
and the Equilibrium Expectation algorithm was demonstrated on an
undirected online social network with 104~103 nodes \cite{byshkin18}.
Using the implementation described in this paper,
modified to use a simplified EE algorithm, Borisenko, Byshkin, \& Lomi
\cite{borisenko19} are able to estimate a simple ERGM model of a
75~879 node directed network.

We note that social networks are typically sparse, and we assume
sparsity for efficient data structures. There is some specific work on
sampling methods for ERGMs for large sufficiently dense networks with
additional assumptions \cite{thiemichen17} such as the presence of
block structure \cite{babkin17,schweinberger18} but here we assume
only sparsity, and that the network can plausibly be described by an
exponential random graph model.

In this paper, we describe an implementation of the EE algorithm,
including the improved fixed density ERGM sampler \cite{byshkin16} for
application to directed networks. By implementing these algorithms and
the associated computations of change statistics in a more efficient
and scalable manner, we are able to estimate ERGM parameters for
networks far larger than previously possible, even using existing
implementations of the algorithms used for the computational experiments
in the papers originally describing them
\cite{byshkin16,byshkin18}. The implementation we describe allows
ERGM parameter estimation for a model of a directed network with over
one million nodes, while existing methods are only practical on
networks of a few thousand nodes at most. We test the implementation
first on simulated networks with known model parameters, in order to
validate that it works correctly, and then demonstrate its application
to an online social network with over 1.6 million nodes.

\subsection*{Exponential random graph models}

An ERGM is a probability distribution with the form

\begin{equation}
  \Pr(X =x) = \frac{1}{\kappa}\exp\left(\sum_A \theta_A z_A(x)\right)
\end{equation}
where
\begin{itemize}
\item $X = [X_{ij}]$ is a 0-1 matrix of random tie variables,
\item $x$ is a realization of $X$,
\item $A$ is a \firstuse{configuration}, a (small) set of nodes and a subset of ties between them,
\item $z_A(x)$ is the network statistic for configuration $A$,
\item $\theta_A$ is a model parameter corresponding to configuration $A$,
\item $\kappa$ is a normalizing constant to ensure a proper distribution.
\end{itemize}

Which configurations $A$ are allowed depends on the assumptions as to
which ties are independent. Here we will use the \firstuse{social
  circuit dependence} assumption \cite{snijders06,robins07},
under which two potential ties are conditionally dependent exactly
when, if they were observed, they would form a 4-cycle in the
network \cite{robins09}. Configurations allowed by other, simpler,
dependence assumptions (Bernoulli, dyad-independent, Markov
\cite[pp.~56--57]{lusher13}) are also allowed in these models.

Under this assumption, we will now describe the structural
configurations used in this work. In the following, $N$ is the number of nodes
in the network.

The simplest configuration, included in every model, is Arc, analogous
to the intercept in a regression. Arc is included to account for the overall density of the
network observed. Its corresponding statistic is $z_L = \sum_{i=1}^{N} \sum_{j=1}^{N} x_{ij}$,
the number of arcs in the graph. The Reciprocity parameter is used to
test for propensity of arcs to be reciprocated, and its statistic
is $z_{\mathrm{Reciprocity}} = \sum_{i=1}^{N} \sum_{j=1}^{N} x_{ij} x_{ji}$.

\begin{figure}[!h]
  \caption{{\bf Alternating $k$-star structures for modeling degree
    distribution in directed networks.}
    Alternating $k$-in-star models
    popularity spread and alternating $k$-out-star models activity
    spread.}
  \label{fig:kstars}
  \includegraphics[scale=0.30]{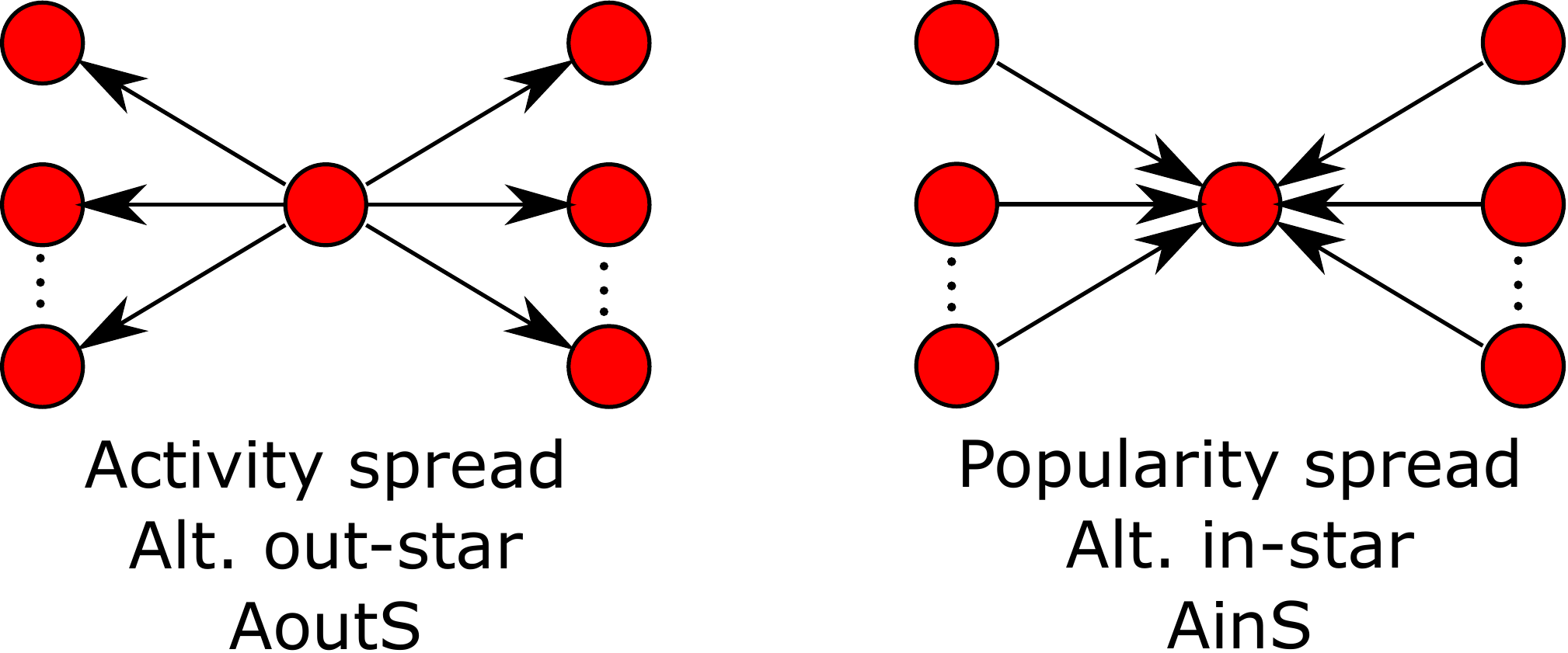} 
\end{figure}

The degree distribution in a
directed network is modeled with the alternating $k$-out-star
and alternating $k$-in-star configurations defined by
\cite{snijders06} and illustrated in
Fig~\ref{fig:kstars}.
The statistic for $k$-out-star is defined as:
\begin{equation}
  z_{\mathrm{AoutS}} = \sum_{k=2}^{N-1}(-1)^k\frac{S_{k}^{Out}}{\lambda^{k-2}}
\end{equation}
where $S_{k}^{Out}$ is the number of
$k$-out-stars and $\lambda \ge 1$ is a damping parameter. We use
$\lambda = 2$ in this work, as used previously in, for example,
\cite{snijders06,stivala16}. We note that in a more general form of
ERGM, the curved exponential family random graph model
\cite{hunter06}, it is also possible to estimate (a parameter
equivalent to) the parameter $\lambda$, and this is routinely done
using the statnet software \cite{handcock08,hunter2008ergm,statnet,ergm}.
However the EE algorithm requires that every model parameter has a
corresponding change statistic, and so cannot estimate curved ERGMs
\cite{byshkin18}. For this reason assume a fixed value for $\lambda$.

The $z_{\mathrm{AinS}}$ statistic for $k$-in-stars is defined
similarly.

\begin{figure}[!h]
  \caption{{\bf Alternating transitive $k$-triangle and alternating
      $k$-2-paths structures.}
    These are used for modeling social circuit
    dependence including path closure and shared popularity. The
    A2P-TD configuration counts $k$-2-paths (A2P-T) and shared
    popularity (A2P-D) configurations in a single configuration,
    adjusting for double counting.}
  \label{fig:tri_and_2paths}
    \includegraphics[scale=0.30]{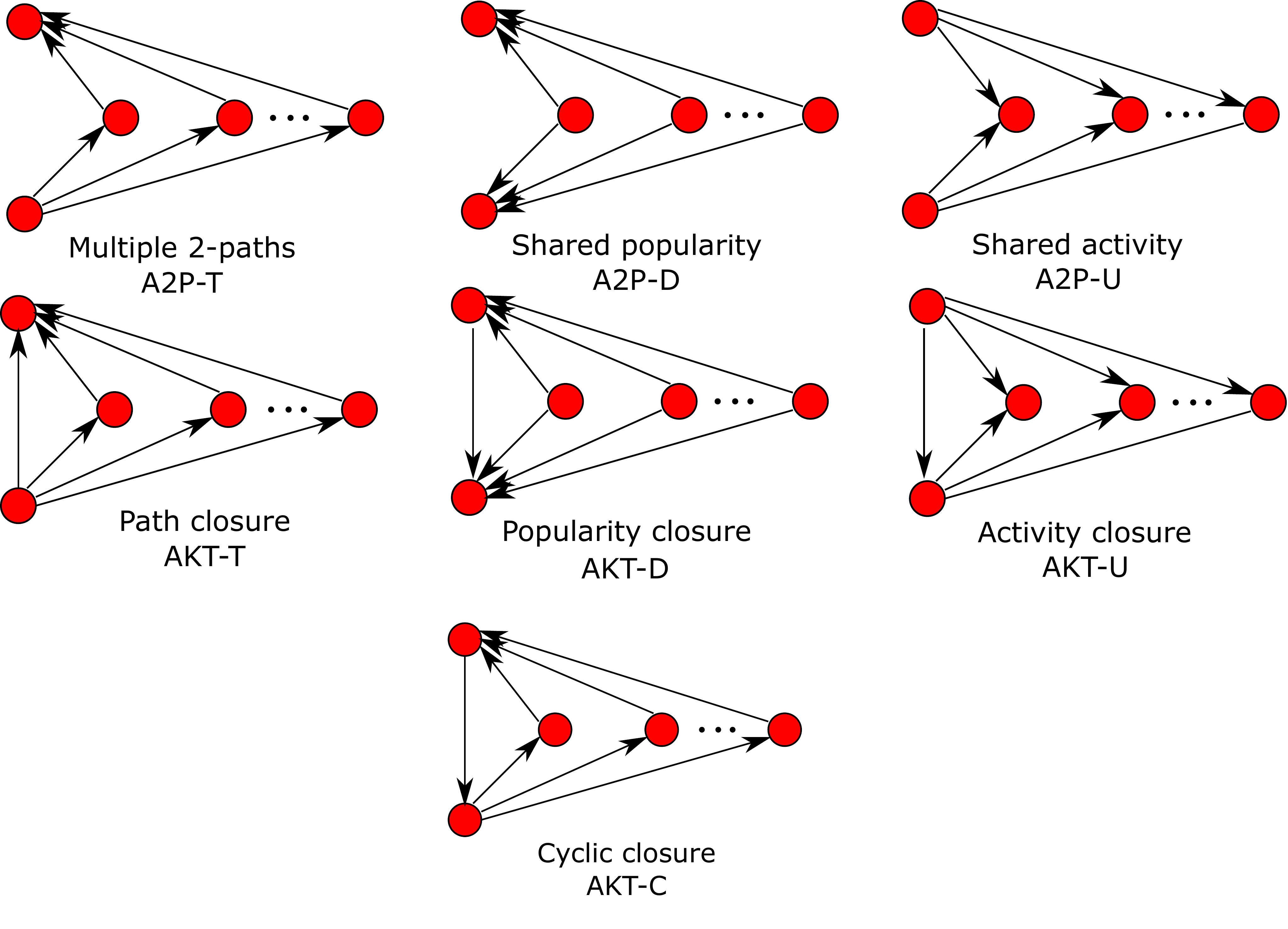} 
\end{figure}

Path closure and multiple connectivity are modeled with the
alternating transitive $k$-triangles and alternating two-paths effects defined by
\cite{snijders06} and illustrated in
Fig~\ref{fig:tri_and_2paths}. These statistics are defined as \cite{wang12thesis}:
\begin{equation}
  z_{\mathrm{AT-T}} = \lambda \sum_{i=1}^N \sum_{j=1}^N x_{ij} \left[ 1 - \left( 1 - \frac{1}{\lambda}\right)^{L_2(i,j)} \right]
\end{equation}
where $L_2(i,j) = \sum_{h=1}^N x_{ih} x_{hj}$ is the number of directed two-paths from $i$ to $j$, and
\begin{equation}
  z_{\mathrm{A2P-T}} = \lambda \sum_{i<j}^N \left[ 1 - \left( 1 - \frac{1}{\lambda} \right)^{L_2(i,j)} \right]
\end{equation}

As well as path closure (AT-T) we can also define cyclic closure
(AT-C), in which arcs constituting the triangles form a cycle. Its
statistic $z_{\mathrm{AT-C}}$ is defined analogously to
$z_{\mathrm{AT-T}}$ but counting cyclic $k$-triangles rather than
transitive $k$-triangles:
\begin{equation}
  z_{\mathrm{AT-C}} = \lambda \sum_{i=1}^N \sum_{j=1}^N x_{ji} \left[ 1 - \left( 1 - \frac{1}{\lambda}\right)^{L_2(i,j)} \right]
\end{equation}
      
We also include the \firstuse{shared popularity} configuration
A2P-D \cite{robins09}, the statistic for which $z_{\mathrm{A2P-D}}$ is
defined analogously to $z_{\mathrm{A2P-T}}$, but rather than counting
directed paths between two nodes via $k$ intermediate nodes,
it counts ``paths'' where the each of the $k$ intermediate nodes
have arcs directed towards each of the two nodes (see
Fig~\ref{fig:tri_and_2paths}):
\begin{equation}
  z_{\mathrm{A2P-D}} = \lambda \sum_{i<j}^N \left[ 1 - \left( 1 - \frac{1}{\lambda} \right)^{L_{2D}(i,j)} \right]
\end{equation}
where $L_{2D}(i,j) = \sum_{h=1}^N x_{hi} x_{hj}$.
We then define the configuration A2P-TD which is the sum of the A2P-D and A2P-T statistics, adjusting for double-counting:
\begin{equation}
  z_{\mathrm{A2P-TD}} =   z_{\mathrm{A2P-T}} + \frac{z_{\mathrm{A2P-D}}}{2}
\end{equation}

The \firstuse{shared activity} configuration A2P-U \cite{robins09},
the statistic for which is $z_{\mathrm{A2P-U}}$ is similar to
A2P-D, but counts ``paths'' where each of the $k$ intermediate nodes
have arcs directed from the pairs of nodes to the intermediate nodes
(see
Fig~\ref{fig:tri_and_2paths}):
\begin{equation}
  z_{\mathrm{A2P-U}} = \lambda \sum_{i<j}^N \left[ 1 - \left( 1 - \frac{1}{\lambda} \right)^{L_{2U}(i,j)} \right]
\end{equation}
where $L_{2U}(i,j) = \sum_{h=1}^N x_{ih} x_{jh}$.

The statistics for the closures corresponding to the open path types
A2P-D and A2P-U, popularity closure (AKT-D) and activity closure
(AKT-U), respectively, are defined similarly to the way path closure
(AKT-T) is defined for the corresponding multiple two-paths A2P-T.

These configurations are illustrated in Fig~\ref{fig:tri_and_2paths}.

\begin{figure}[!h]
  \caption{{\bf Binary attribute configurations.}
    The dark nodes represent
    actors with the binary attribute, and the lighter shaded nodes
    represent actors with or without the attribute.}
  \label{fig:binary_configurations}
    \includegraphics[scale=0.30]{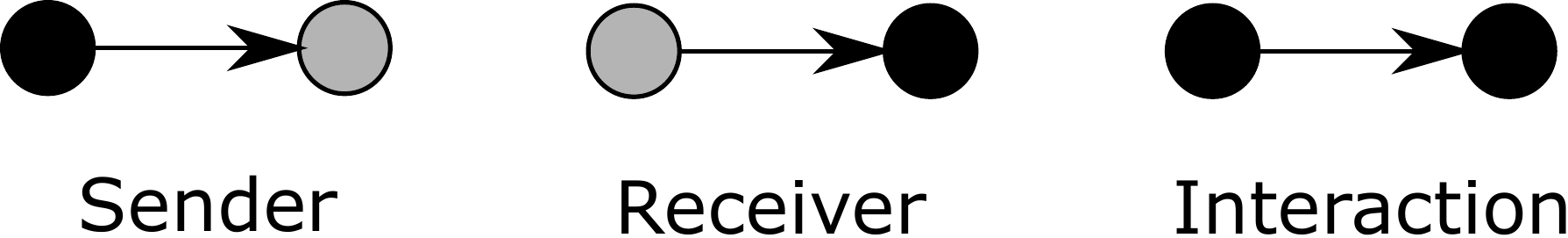} 
\end{figure}

In addition, we will allow nodes to have binary, categorical, or continuous
attributes, and use the following additional configurations using
these nodal attributes. For the binary attribute, we use the
four configurations \firstuse{Sender}, \firstuse{Receiver}, and
\firstuse{Interaction}, illustrated in
Fig~\ref{fig:binary_configurations}.
The Sender parameter indicates increased propensity of a node with the True
value of the binary attribute to ``send'' a tie to another node, and the
Receiver the increased propensity of a node with the True value of the
attribute to ``receive'' a tie from another node (both irrespective
of the attribute value of the other node). The Interaction parameter
indicates increased propensity for two nodes both with the True value
of the attribute to have an arc connecting them. The corresponding
statistics are defined as follows (where we now use the notation $\sum_{i,j}$ for summation over all pairs of nodes $i \in \left\{1 \ldots N \right\}$, $j \in \left\{1 \ldots N \right\}$, $i \neq j$):
\begin{eqnarray}
  z_{\mathrm{Sender}} & = & \sum_{i,j} a_i x_{ij} \\
  z_{\mathrm{Receiver}} &=& \sum_{i,j} a_j x_{ij} \\
  z_{\mathrm{Interaction}} &=& \sum_{i,j} a_i a_j x_{ij} 
\end{eqnarray}
where $a_i \in \left\{0,1\right\}$ is the value of the binary
attribute on node $i$.

\begin{figure}[!h]
  \caption{{\bf Categorical attribute configurations.}
    The filled and empty
    nodes represent actors with two different values of the categorical
    attribute.}
  \label{fig:categorical_configurations}
    \includegraphics[scale=0.30]{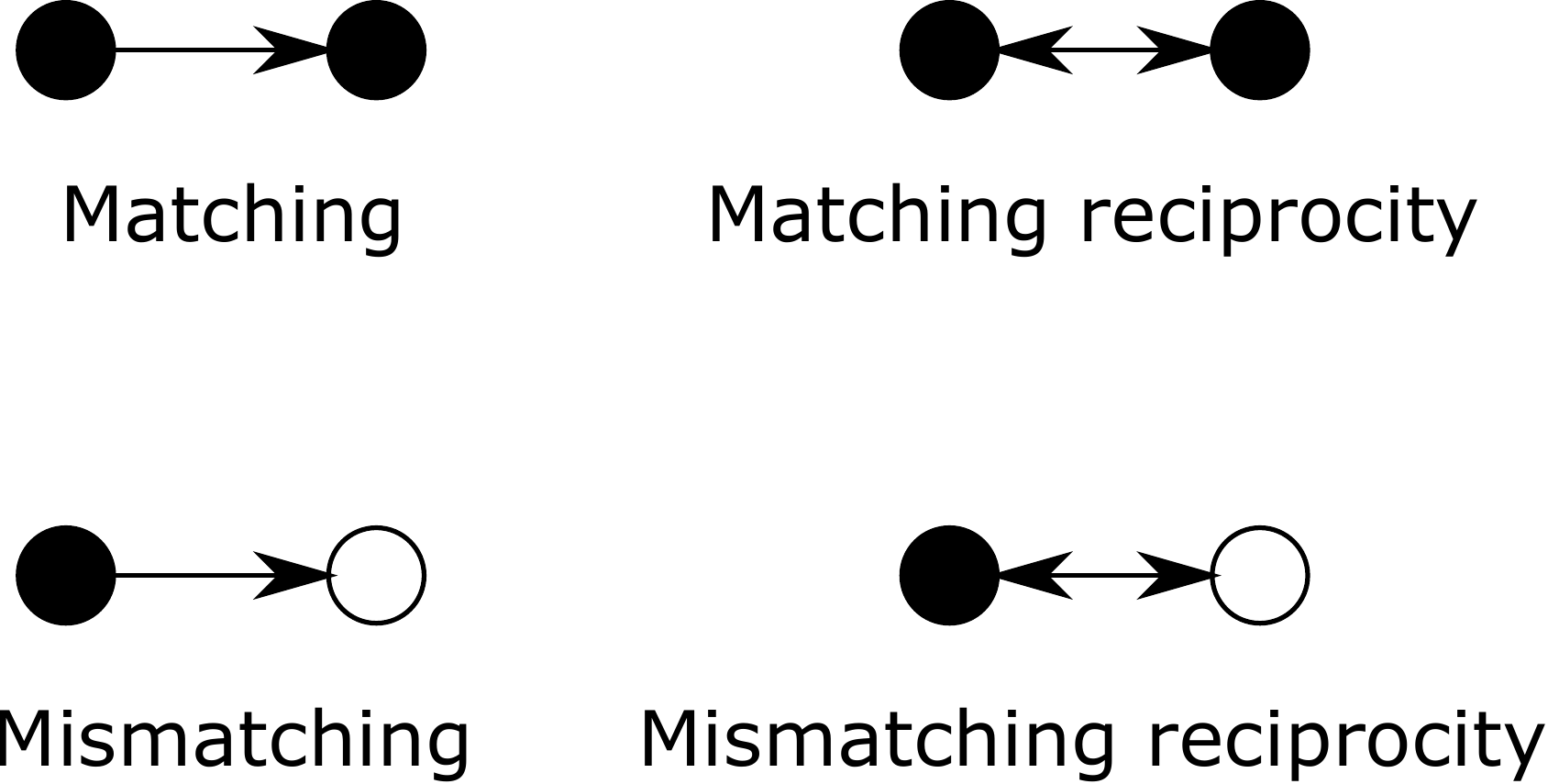} 
\end{figure}

For the categorical attribute, the \firstuse{Matching} and
\firstuse{Mismatching} parameters indicate the
increased propensity for a node to send a tie to another node with,
respectively, the same, or different, value of the categorical
attribute. The \firstuse{Matching reciprocity} and
\firstuse{Mismatching reciprocity} parameters indicate the
increased propensity for such ties to be reciprocated.
These configurations are illustrated in
Fig~\ref{fig:categorical_configurations} and the corresponding
statistics are defined by:
\begin{eqnarray}
  z_{\mathrm{Matching}} &=& \sum_{i,j} \delta_{c_i, c_j} x_{ij} \\
  z_{\mathrm{Mismatching}} &=&  \sum_{i,j} (1-\delta_{c_i, c_j}) x_{ij} \\
  z_{\mathrm{Matching~reciprocity}} &=& \sum_{i,j} \delta_{c_i, c_j} x_{ij} x_{ji} \\
  z_{\mathrm{Mismatching~reciprocity}} &=& \sum_{i,j}  (1-\delta_{c_i, c_j}) x_{ij} x_{ji} 
\end{eqnarray}
where $c_i$ is the value of the categorical attribute at node $i$ and
$\delta_{x,y}$ is the Kronecker delta function
\begin{equation}
  \delta_{x,y} = \left\{
  \begin{array}{lll}
    0 & \mathrm{if} & x \neq y, \\
    1 & \mathrm{if} & x = y. 
  \end{array}
  \right.
\end{equation}

For a continuous attribute $u_i$ on a node $i$, we also define the
(continuous) Sender, Receiver and Difference statistics as follows:
\begin{eqnarray}
  z_{\mathrm{continuousSender}} & = & \sum_{i,j} u_i x_{ij} \\
  z_{\mathrm{continuousReceiver}} &=& \sum_{i,j} u_j x_{ij} \\
  z_{\mathrm{Diff}} &=& \sum_{i,j} \abs{u_i - u_j} x_{ij} 
\end{eqnarray}
These indicate, respectively, the increased propensity of a node to send
ties for higher values of its continuous  attribute, the increased propensity
of a node to receive ties for higher values of its continuous attribute,
and the increased propensity of nodes to have a tie between them for smaller
absolute differences in their continuous attributes. The latter is a simple
measure of homophily, the tendency for nodes with similar values of the
attribute to have a tie between them.

Note that in ERGM estimation algorithms these statistics as defined
above never actually need to be computed directly. Instead only the
corresponding \emph{change statistics} are computed
\cite{snijders06,hunter06,hunter08,hunter2008ergm,hunter12}. The
change statistic is the change in the statistic due to the addition or
deletion of an arc, which is much faster to compute. For example the
most basic statistic is $z_L$, the count of the number of arcs in the
graph.
Computing this statistic therefore requires counting the number of
arcs in the graph, however the corresponding change statistic is
simply the constant 1 (or -1 for deleting an arc): adding an arc
increases the statistic by 1, and deleting an arc decreases it by 1.

\subsection*{Equilibrium expectation algorithm}

Monte Carlo based methods, such as Markov chain Monte Carlo
maximum likelihood estimation (MCMCMLE) \cite{hunter06} and stochastic
approximation \cite{snijders02} as well as Bayesian methods
\cite{caimo11}, as reviewed for example by Hunter \etal
\cite{hunter12}, all require drawing simulated networks from the ERGM
distribution. This can be achieved using a Metropolis--Hastings
algorithm, and a number of samplers are available
\cite{snijders02,morris08,hunter12,byshkin16}.  However all these
methods require that a number of network samples are drawn from the
stationary ERGM distribution, for each updated value of the parameter
vector being estimated, which may require a very large number of
iterations, and limits the size of networks to which these methods can
be applied in practice.

In contrast, the EE algorithm \cite{byshkin18} does not require these
potentially very long MCMC simulations between parameter updates. The
EE algorithm is related to persistent contrastive divergence (PCD)
\cite{younes88,tieleman08,borisenko19} and is fast because it adjusts
its parameters according the difference between the observed network
statistics and the statistics of a current non-equilibrium state of
the Markov chain of simulated networks. It may be thought of as a kind
of gradient ascent method, and depends on the property of the
exponential family (to which the ERGM distribution belongs) that the
expected value of a statistic is a monotonically increasing function
of the corresponding parameter \cite[Ch.~8]{barndorff14}. It works by starting the chain of
simulated networks at the observed network (not the empty network for
example), and taking only a small number of Metropolis--Hastings steps,
before adjusting the estimated parameter values according to the
divergence of the simulated network statistics from the observed
network statistics. After sufficiently many iterations of this process
(which in practice is many orders of magnitude smaller than the number
of Metropolis--Hastings steps required to find the stationary ERGM
distribution), the divergence of each of the statistics from the
observed statistics oscillates around zero, and the corresponding
parameters oscillate around a value which is taken to be an estimate
of the MLE.

A version of contrastive divergence (CD) \cite{hinton02} is used to
compute initial values of the ERGM parameter estimates
\cite{asuncion10,krivitsky17} for the EE algorithm \cite{byshkin18}.
Details of the EE algorithm, as first described in \cite[Supplementary
  Information]{byshkin18}, and of the IFD sampler, \cite{byshkin16}
are provided in \nameref{S1_Appendix}.

\section*{Materials and methods}

\subsection*{Parameter estimation}

Parameters are estimated using a new implementation of the EE
algorithm, which we call EstimNetDirected. This implementation has
change statistics for directed (rather than undirected as in the
original description \cite{byshkin18}) networks, and uses efficient
data structures in order to scale to very large (over one million
node) networks. Both the ``basic'' ERGM sampler (as used in the PNet
\cite{wang09} software) and the improved fixed density (IFD) ERGM
sampler \cite{byshkin16} are implemented.

The network is stored as an adjacency list data structure for space
efficiency and fast computation of change statistics.  A ``reversed''
adjacency list is also maintained.  This stores, for each node $j$, a
list of nodes $i$ for which the arc $i \rightarrow j$ exists. This
allows efficient computation of change statistics that require the
in-neighbours of a node. In addition, a flat list of all arcs is
maintained, for efficient implementation of the IFD sampler, which
requires finding an arc uniformly at random for arc deletion
moves \cite{byshkin16}.

For efficient computation of the ``alternating'' two-path and triangle
change statistics, it is necessary to keep track of the counts of
two-paths between each pair of nodes. It is not scalable to store
these two-paths matrices as arrays as in earlier implementations
\cite{wang09,byshkin18}, so instead hash tables can be used, where the
key is a node pair $(i,j)$ and the value is the relevant two-path
count (which is zero if the key is not present). This takes advantage
of the sparsity (approximately 0.06\% nonzero in the empirical example
described here) of these matrices and still allows fast
(asymptotically constant time) lookup. In addition, a Bloom filter
\cite{bloom70} is used so that the overwhelmingly more frequent case
of looking up an entry that is not present is faster. During the MCMC
ERGM sampling process, in which arcs are added and deleted, entries in
the two-path tables that fall to zero are deleted, in order to stop
the tables from growing in size indefinitely, however this diminishes
the effectiveness of the Bloom filter.

We run a number of estimations independently (and in parallel to
minimize elapsed time).

\subsection*{Standard error estimation}

For each parallel estimation run, the point estimates and their
standard errors are estimated. The point estimate (mean) and
asymptotic covariance matrix for MCMC standard error are estimated
using the multivariate batch means method \cite{jones06,vats17} using
the \unix{mcmcse} R package \cite{mcmcse}. The covariance matrix for
the error in inherent in using the MLE is estimated as the inverse of
the covariance matrix of the simulated statistics (Fisher information
matrix) \cite{snijders02,hunter06} also using \unix{mcmcse}. The total
estimated covariance matrix is then estimated as the sum of these two
covariance matrices, and from this we compute the standard error as
the square root of the diagonal.

The overall estimate and its standard error are then computed as the
inverse variance weighted average \cite[Ch.~4]{hartung08} of the
results calculated from each independent (parallel) estimation.

\subsection*{Implementation and availability}

EstimNetDirected is implemented in the C programming language using
the message passing interface (MPI) for parallelization on computing
clusters. The \unix{uthash} \cite{uthash} macro collection is used to
implement hash tables (including Bloom filter) and the Random123
counter-based pseudo-random number generator \cite{salmon11} is used
to generate pseudo-random numbers for the MCMC process.  Scripts for
processing network data formats, estimating standard errors, and
generating plots and fitting heavy-tailed distributions are written in
R and Python and use the \unix{igraph} \cite{csardi06}, \unix{SNAP}
\cite{leskovec16}, \unix{ggplot2} \cite{wickham09}, \unix{PropCIs}
\cite{propcis}, \unix{mcmcse} \cite{mcmcse}, and \unix{poweRlaw}
\cite{gillespie15} packages.

All source code and scripts are publicly available on GitHub at
\url{https://github.com/stivalaa/EstimNetDirected}.

\subsection*{Simulated networks}

To ensure that the parameter estimation algorithm works correctly,
we first apply it to estimating ERGM parameters of networks with known
true values, and measure the bias, root mean square error (RMSE), coverage, and
Type I and Type II error rates in statistical inference. To do so we
simulate sets of 100 networks sampled from an ERGM network
distribution with known parameters, and then estimate the parameters
of each of the 100 networks with EstimNetDirected. This allows the
mean bias and RMSE to be estimated.
The coverage is then the percentage of the 95\% confidence
intervals which contain the true value of the parameter. Coverage
higher than the nominal 95\% indicates overly conservative (high) estimates of
the standard error, and coverage lower than the nominal value indicates
overly optimistic (low) values of the standard error (uncertainty).
In addition, we estimate the Type
II error rate in inference (the false negative rate), as the
percentage of estimations in which 
the estimated 95\% confidence interval includes zero.

To estimate the Type I error rate (false positive rate) for inference
of an ERGM parameter significance, we generate simulated networks in
which the parameter in question is zero, and proceed as just
described. Then the Type I error rate is estimated as the percentage
of estimations in which the 95\% confidence interval does not include
zero.

We generate two sets of graphs from ERGM distributions, both
with $N = 2000$ nodes. First, a network with binary node attributes
and parameters (Arc, Reciprocity, AinS, AoutS, AT-T, A2P-TD, Interaction,
Sender, Receiver) =  (-1.00, 4.25, -2.00, -1.50, 0.60, -0.15, 2.00, 1.50, 1.00),
and, second, a network with categorical node attributes and parameters
(Arc, Reciprocity, AinS, AoutS, AT-T, A2P-TD, Matching, Matching reciprocity) =
(-1.00, 4.25, -2.00, -1.50, 1.00, -0.15, 1.50, 2.00).
For each of the two sets of parameters, we generated 100 samples from
a network distribution with those parameters using PNet
\cite{wang09}. For networks with a binary attribute, 50 of the nodes
(2.5\%), selected at random, have the True value, and the rest
False. For networks with a categorical attribute, the attribute at
each node is assigned one of three possible values uniformly at
random.  The networks are sampled with sufficient burn-in (of the
order of $10^9$ iterations) to ensure initialization effects are
minimized, and samples are taken sufficiently far apart (separation of
the order of $10^8$ iterations) to ensure that they are essentially
independent.  Table~\ref{tab:graphstats} shows summary statistics of
the simulated networks.

\begin{table}[!ht]
\begin{adjustwidth}{-2.25in}{0in} 
  \centering
  \caption{{\bf Statistics of the simulated directed networks.}}
\begin{tabular}{|r|l|l|r|r|r|r|}
\hline
N & Attributes & Zero effect  & Mean components & Mean degree  & Mean density  & Mean c.c.  \\
\thickhline
2000 & Binary & None & 1.05 & 2.65 & 0.00132 & 0.00482\\ \hline
2000 & Categorical & None & 1.00 & 3.79 & 0.00189 & 0.05655\\ \hline
2000 & Binary & Sender & 1.02 & 2.60 & 0.00130 & 0.00323\\ \hline
2000 & Binary & Receiver & 1.03 & 2.61 & 0.00131 & 0.00338\\ \hline
2000 & Binary & Reciprocity & 1.03 & 2.58 & 0.00129 & 0.00339\\ \hline
2000 & Binary & Interaction & 1.00 & 2.63 & 0.00131 & 0.00326\\ \hline
2000 & Binary & AT-T & 1.00 & 2.63 & 0.00132 & 0.00215\\ \hline
2000 & Binary & A2P-TD & 1.00 & 3.68 & 0.00184 & 0.13192\\ \hline
2000 & Binary & AinS & 1.00 & 9.95 & 0.00498 & 0.32332\\ \hline
2000 & Binary & AoutS & 1.00 & 6.74 & 0.00337 & 0.36730\\ \hline
2000 & Categorical & Matching & 1.04 & 2.67 & 0.00134 & 0.00643\\ \hline
2000 & Categorical & Match. Recip. & 1.00 & 3.16 & 0.00158 & 0.00966\\ \hline
2000 & Categorical & Reciprocity & 1.00 & 3.04 & 0.00152 & 0.00707\\ \hline
2000 & Categorical & AT-T & 1.00 & 3.49 & 0.00175 & 0.00331\\ \hline
2000 & Categorical & A2P-TD & 5.97 & 1.83 & 0.00092 & 0.00802\\ \hline
2000 & Categorical & AinS & 1.00 & 5.38 & 0.00269 & 0.10851\\ \hline
2000 & Categorical & AoutS & 1.11 & 3.16 & 0.00158 & 0.04159\\ \hline
\end{tabular}
\begin{flushleft}
``c.c.'' is the global clustering coefficient.
Note that for the Categorical attribute networks with AinS,
AoutS, and A2P zero effects, the Arc parameter is set to -4.0 
rather than -1.0 to avoid the network becoming dense.
\end{flushleft}
\label{tab:graphstats}
\end{adjustwidth}
\end{table}

\subsection*{Empirical network}

As an example application we use EstimNetDirected to estimate an ERGM
model of the Pokec online social network \cite{takac12} which is
publicly available from the Stanford large network dataset collection
\cite{snapnets}. Pokec is the most popular online social network in
Slovakia \cite{takac12} and represents a sizable percentage of
Slovakia's population \cite{kleineberg14}. This publicly available
data is also unusual and particularly useful as a test case for social
network algorithms as it is the entire online social network at a point
in time, rather than a sample as is often the case, and the
nodes are annotated with attributes, specifically including gender, age, and
region (187 of them, Slovakia or elsewhere) which we use as nodal
covariates in the model.

This network has 1 632 803 nodes and 30 622 564 arcs (directed graph
density approximately $10^{-5}$ and mean degree 37.5).  The nodes are
users of the Pokec online social network, and arcs represent directed
``friendship'' relations, \ie unlike many online social networks,
``friendships'' are not assumed to be automatically reciprocated
(undirected). In fact only approximately 54\% of the ``friendship''
relations are reciprocated.  More details and descriptive statistics
of this network are in \cite{takac12}.

The Pokec online social network has been previously used as a test bed
for social network analysis algorithms, and in particular by
Kleineberg \& Bogu{\~n}{\'a} \cite{kleineberg14} who use it to test a
model of the evolution of an online social network. They treat the
final state of the network as a representation of a true social
network, as do we, in order to demonstrate EstimNetDirected estimation of an
ERGM, which is a cross-sectional network model. However they treat
the network as undirected, including only reciprocated ties, while we
maintain the directed nature of the network.

\begin{figure}[!h]
  \caption{{\bf Pokec network degree distribution.}  Neither the in-
    nor the out-degree distribution are consistent with power law or
    log-normal distributions ($p < 0.01$).}
  \label{fig:pokec_powerlaw}
    \includegraphics[scale=0.8]{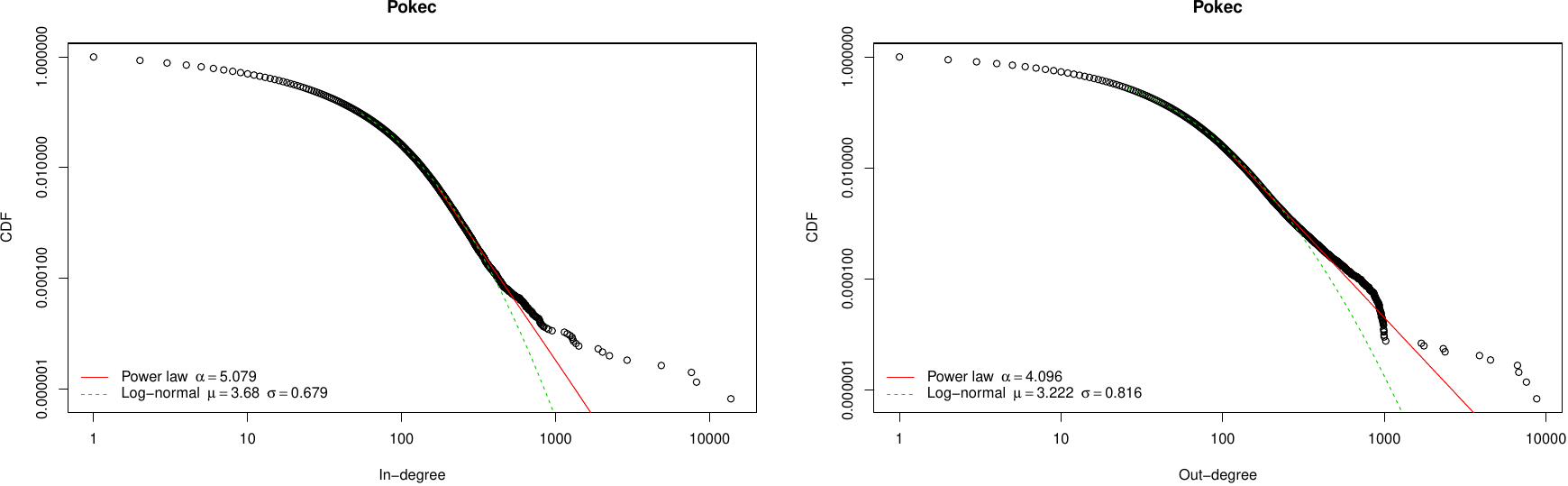} 
\end{figure}

The Pokec degree distribution was described as ``scale-free''
in \cite{takac12}, but based only on visual examination of the
degree-frequency plot. This technique, or similarly, fitting a straight
line to a degree-frequency log-log plot, is now well known to be
not a sound technique for assessing whether a distribution is scale-free
or follows a power law \cite{clauset09,stumpf12,broido19}.
Using the statistical method described by Clauset \etal \cite{clauset09}
as implemented in the powerRlaw R package \cite{gillespie15},
we find that the neither the in- nor out-degree distribution of
the Pokec online social network is consistent with a power
law distribution (Fig~\ref{fig:pokec_powerlaw}).

Nevertheless, it is clear from Fig~\ref{fig:pokec_powerlaw} that
there are \firstuse{hubs} in the network, that is, nodes with an order
of magnitude higher degree than most other nodes. In particular, there
is a noticeable ``break'' in the empirical cumulative distribution
function (CDF) plot at degree 1000, most noticeable for the out-degree
distribution. According to Takac \& Zabovsky, ``hubs in Pokec are not
people but commercial companies...'' \cite[p.~5]{takac12}.

\begin{figure}[!h]
  \caption{{\bf Pokec network degree distribution after hub nodes are
      removed.}  After removing the 20 nodes that have in- or
    out-degree greater than 1000, the resulting network's in-degree
    distribution is consistent with a power law distribution, but not
    with a log-normal distribution ($p<0.01)$. The out-degree
    distribution is consistent with neither a power law nor a
    log-normal distribution ($p<0.01$).}
  \label{fig:pokec_nohubs_powerlaw}
    \includegraphics[scale=0.8]{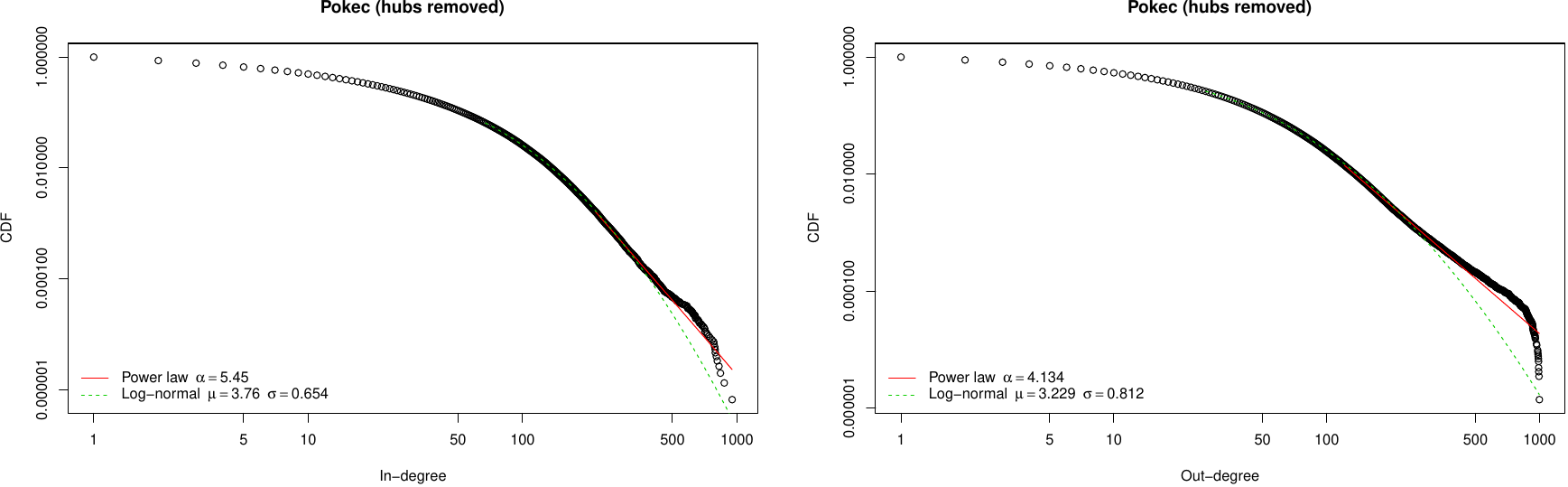} 
\end{figure}

Based on these observations, and on the fact that an initial attempts
to estimate an ERGM for the entire network did not converge, we remove
all nodes with in- or out-degree greater than 1000 from the
network. There are only 20 such nodes (0.001\% of the nodes), and
removing them does not significantly change the density or mean
degree. However it breaks the network, which was initially a single
connected component, into 577 components, although the giant component
has 1 632 199 nodes (99.96\% of the total).  This indicates that these
hubs are not performing the function of holding different components
of the network together into a connected whole, as their removal only
results in the creation of a relatively small number of isolated
nodes, rather than splitting the network into multiple large
components, and that therefore their removal is not substantively
changing the nature of the network structure. The in-degree
distribution of network with hubs removed is consistent with a power
law, although the out-degree distribution is not
(Fig~\ref{fig:pokec_nohubs_powerlaw}).

The EstimNetDirected parameter settings used in the estimations
are detailed in \nameref{S1_Table}.

\subsection*{Convergence tests}

As in \cite{byshkin18} we use a t-ratio check for convergence, but for
larger directed networks we weaken the criterion to conclude non-convergence
if the absolute value of any parameter's t-ratio is greater than 0.3.
If the covariance matrix computed in the standard error
estimation step is (nearly) computationally singular then that estimate
is considered non-converged, possibly due to model degeneracy.

For empirical network estimations where there are not a large number
of automated estimations to process, visual inspection of the
parameter and statistic trace plots is used as a heuristic to confirm
convergence.  In the case of the simulated networks where large
numbers of estimations are run, we automate the additional heuristic
that estimations with numeric overflow (``NaN'' values) or ``huge''
(greater in magnitude than $10^{10}$) parameter values are
non-converged.

An additional heuristic (visual) convergence test for modeling
empirical networks is to plot various network statistics of the observed
network on the same plot as the distribution of these statistics in
the EE algorithm simulated networks. This is the same principle as the
t-ratio test, but it includes statistics other than those explicitly
included in the model in order to give some indication of how well
the model fits. The statistics included are the degree distribution,
reciprocity, giant component size, average local and global clustering
coefficients, triad census, geodesic distribution, and edgewise and
dyadwise shared partners (similar to goodness-of-fit plots in
statnet \cite{handcock08,hunter2008ergm,statnet,ergm}).

Note, however, that these plots are not goodness-of-fit plots, as the
simulated networks are not generated \latin{ab initio} (\eg from the
empty network) from the estimated parameters, but rather are the
networks that have been simulated in the EE algorithm where the
starting point is the observed network. Hence the plots may be
``over-optimistic'' in indicating a good model fit: however a plot
clearly showing a poor model fit definitely indicates lack of
convergence or a poor model.

In addition, for every large networks it is impractical to compute
some distributions in reasonable time (such as the edgewise and
dyadwise shared partners, and the geodesic distribution) and so
these plots are excluded for very large networks.
An example of a convergence plot is shown
in \nameref{S1_Fig}.

\section*{Results and discussion}

\subsection*{Simulated networks}

\begin{figure}[!h]
  \caption{{\bf EstimNetDirected parameter estimates for 2000 node
      networks with categorical attribute.}
    The error bars show the nominal 95\% confidence interval.
    The horizontal line shows
    the true value of the parameter, and each plot is also annotated
    with the mean bias, root mean square error (RMSE), the percentage
    of samples for which the true value is inside the confidence
    interval, the coverage (\% in CI), and the Type II error rate (False Negative Rate, FNR).
    $\mathrm{N_C}$ is the number of networks (of the total 100) for which a
    converged estimate was found, each of which is shown on the plot.}
  \label{fig:cat_meanse}
  \includegraphics[width=\textwidth]{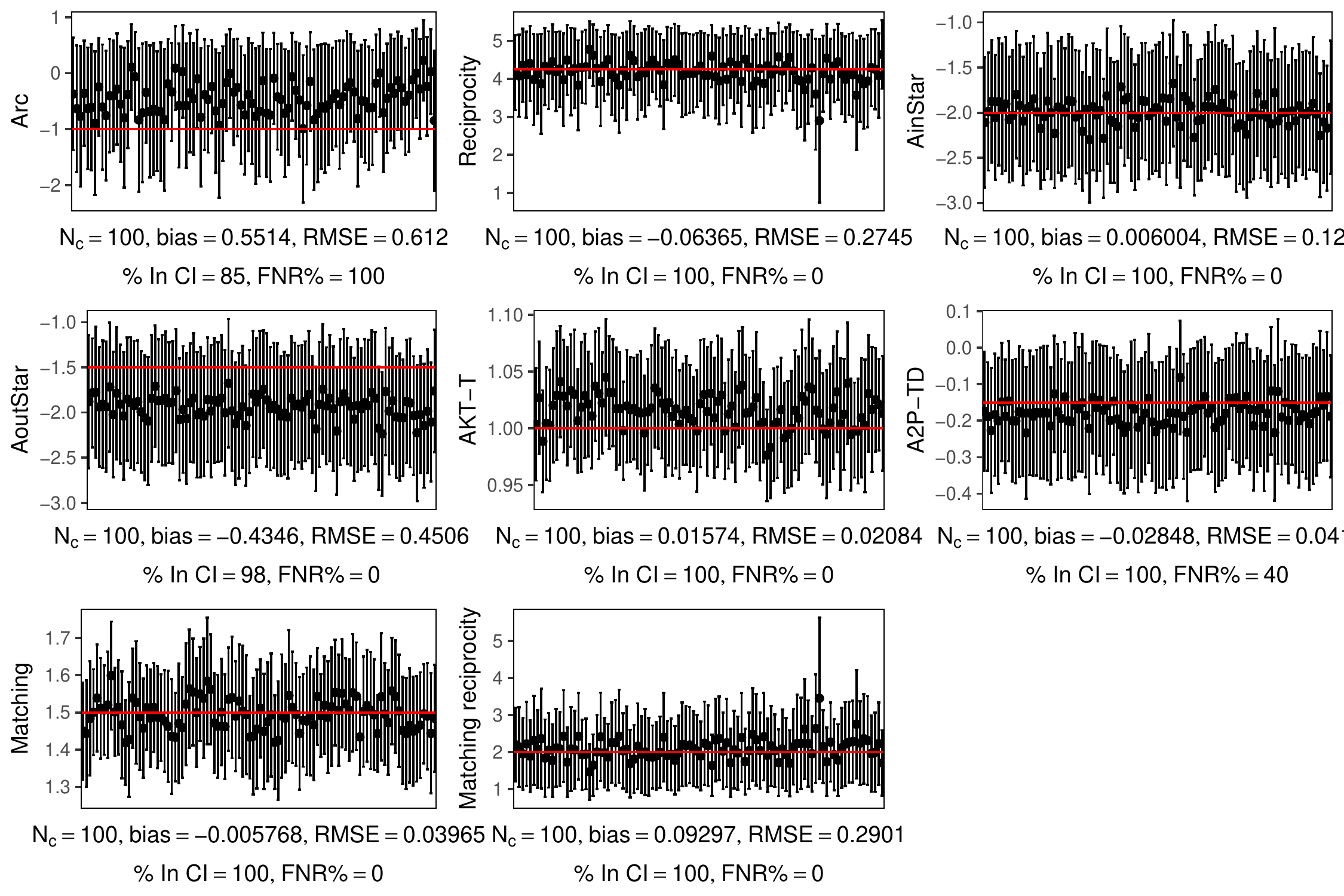} 
\end{figure}

\begin{figure}[!h]
  \caption{{\bf EstimNetDirected parameter estimates for 2000 node
      networks with binary attribute.} The error bars show the nominal 95\% confidence interval. The horizontal line shows the
    true value of the parameter, and each plot is also annotated with
    the mean bias, root mean square error (RMSE), the percentage of
    samples for which the true value is inside the confidence
    interval, the coverage (\% in CI), and the Type II error rate (False Negative Rate, FNR).
    $\mathrm{N_C}$ is the number of networks (of the total 100) for which a
  converged estimate was found, each of which is shown on the plot.}
  \label{fig:bin_meanse}
  \includegraphics[width=\textwidth]{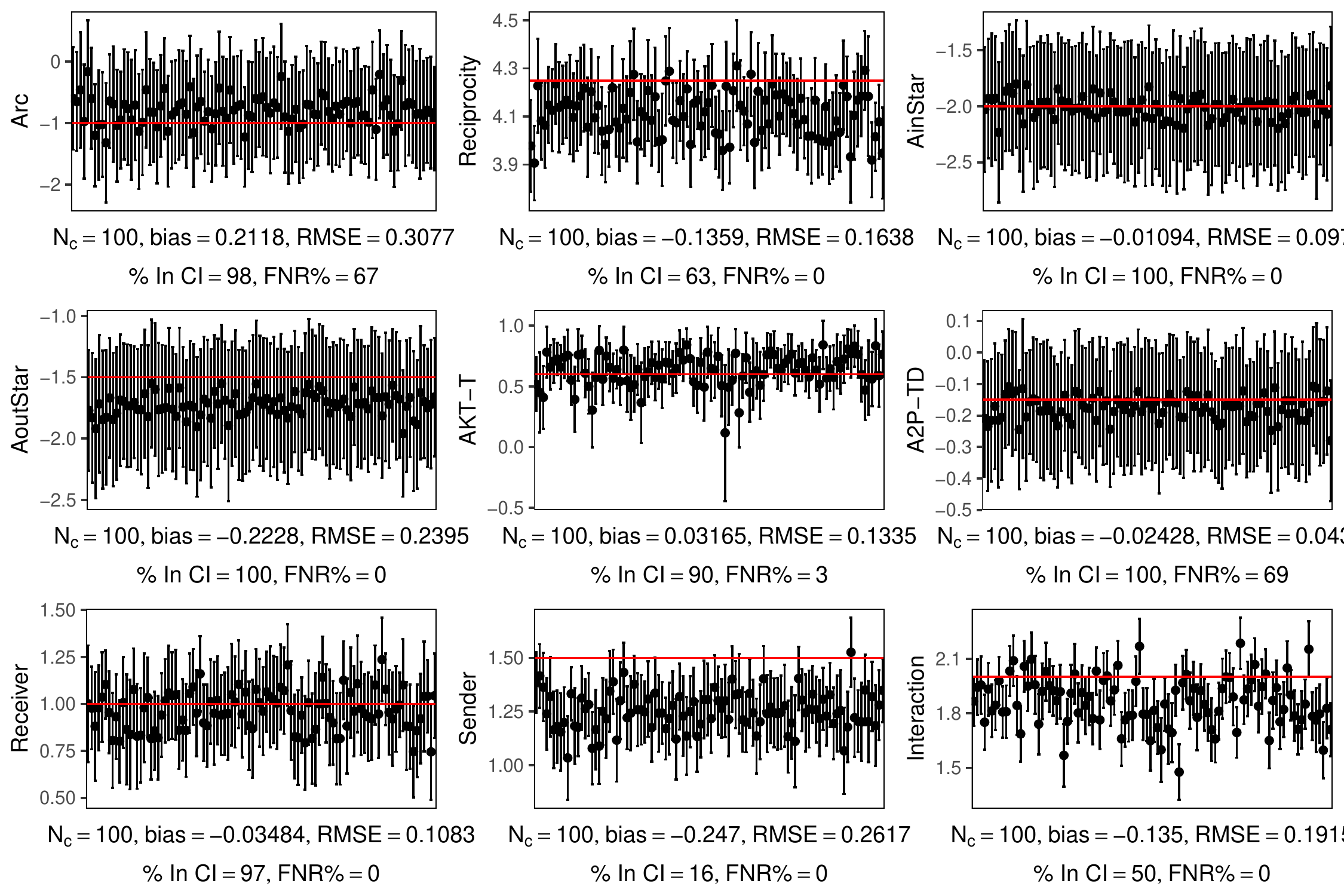} 
\end{figure}

Table~\ref{tab:fnr_rate} shows the bias, RMSE, Type II error rate, and
coverage in estimating the simulated networks. It can be seen that the
Type II error rate in inference is very low on all parameters except
for Arc and and A2P-TD. In the case of the Arc parameter, this is of
little concern as we do not in practice need to make a statistical
inference on this parameter, as previously noted it is analogous to an
intercept and simply used to control for network density.
Fig~\ref{fig:cat_meanse} and Fig~\ref{fig:bin_meanse}, plotting the
point estimates and their error bars, shed some more light on this
problem. It seems the high Type II error rate for Arc and A2P-TD is at
least partly explained by the small magnitude of the true values of
these parameters. The coverage in most of these cases
indicates overly conservative error estimates (higher than
the nominal 95\%), and the Type II error rate is high.

Additionally, in the case of the Arc parameter especially, the
parameter estimates appear to be biased. Although the Type II error
rate is very low, in the case of the networks with binary attribute,
the coverage rate is low for the Sender, Interaction, and (to a lesser
degree) Reciprocity parameters.  Fig~\ref{fig:bin_meanse} shows that
this appears to be because of bias in these parameter estimates. In
the case of the networks with a categorical attribute,
Fig~\ref{fig:cat_meanse} shows positive bias in the Arc parameter
resulting in a high Type II error rate despite lower than nominal coverage.

We should not be surprised by bias in the parameter estimates, as the
MLE for ERGM canonical parameters is biased precisely because it is
unbiased by construction in the mean value parameter space
\cite{vanduijn09} (that is, the mean values of the statistics, not the
corresponding ERGM parameters). For this reason, van Duijn, Gile, \&
Handcock \cite{vanduijn09} propose a framework for assessing estimators
in which bias is compared in the mean value parameter space, by
generating large numbers of simulated networks from the estimated
parameter values, and comparing them to the statistics of the original
simulated networks. However for large directed networks this procedure
is impractical due to the time it takes to simulate each set of
networks (the very reason the EE algorithm is so fast is that it
avoids doing this).  In addition, as the purpose of ERGM parameter
estimation is usually to make statistical inferences from the
estimated parameters, it is useful to measure the inferential error in
this procedure.

\begin{table}
  \begin{adjustwidth}{-2.25in}{0in} 
    \centering
    \caption{{\bf Results from estimation of simulated networks using
        EstimNetDirected estimating Type II error rate.}}
\begin{tabular}{|r|l|l|r|r|r|r|r|r|r|r|}
\hline
N &  Attributes &  Effect &  Bias &  RMSE & estim. & lower  & upper & in C.I. (\%)    & $N_C$     & $\overline{N_R}$      \\
\thickhline
2000  &  Categorical  &   A2P-TD  & -0.0285 & 0.0411 & 40 & 31 & 50 & 100 & 100 & 32.00 \\ \hline
2000  &  Categorical  &   AinS  & 0.0060 & 0.1298 &  0 &  0 &  4 & 100 & 100 & 32.00 \\ \hline
2000  &  Categorical  &   AKT-T  & 0.0157 & 0.0208 &  0 &  0 &  4 & 100 & 100 & 32.00 \\ \hline
2000  &  Categorical  &   AoutS  & -0.4346 & 0.4506 &  0 &  0 &  4 & 98 & 100 & 32.00 \\ \hline
2000  &  Categorical  &   Arc  & 0.5514 & 0.6120 & 100 & 96 & 100 & 85 & 100 & 32.00 \\ \hline
2000  &  Categorical  &   Matching  & -0.0058 & 0.0396 &  0 &  0 &  4 & 100 & 100 & 32.00 \\ \hline
2000  &  Categorical  &   MatchingReciprocity  & 0.0930 & 0.2901 &  0 &  0 &  4 & 100 & 100 & 32.00 \\ \hline
2000  &  Categorical  &   Reciprocity  & -0.0636 & 0.2745 &  0 &  0 &  4 & 100 & 100 & 32.00 \\ \hline
2000  &  Binary  &   A2P-TD  & -0.0243 & 0.0439 & 69 & 59 & 77 & 100 & 100 & 31.98 \\ \hline
2000  &  Binary  &   AinS  & -0.0109 & 0.0974 &  0 &  0 &  4 & 100 & 100 & 31.98 \\ \hline
2000  &  Binary  &   AKT-T  & 0.0316 & 0.1335 &  3 &  1 &  8 & 90 & 100 & 31.98 \\ \hline
2000  &  Binary  &   AoutS  & -0.2228 & 0.2395 &  0 &  0 &  4 & 100 & 100 & 31.98 \\ \hline
2000  &  Binary  &   Arc  & 0.2118 & 0.3077 & 67 & 57 & 75 & 98 & 100 & 31.98 \\ \hline
2000  &  Binary  &   Interaction  & -0.1350 & 0.1915 &  0 &  0 &  4 & 50 & 100 & 31.98 \\ \hline
2000  &  Binary  &   Receiver  & -0.0348 & 0.1083 &  0 &  0 &  4 & 97 & 100 & 31.98 \\ \hline
2000  &  Binary  &   Reciprocity  & -0.1359 & 0.1638 &  0 &  0 &  4 & 63 & 100 & 31.98 \\ \hline
2000  &  Binary  &   Sender  & -0.2470 & 0.2617 &  0 &  0 &  4 & 16 & 100 & 31.98 \\ \hline
\end{tabular}
\begin{flushleft}
  The ``estim.'', ``lower'', and ``upper'' columns show the point
  estimate and lower and upper 95\% confidence interval (C.I.),
  respectively, of the Type II error rate (false negative rate). This
  C.I. is computed as the Wilson score interval \cite{wilson27}.  The
  ``in C.I. (\%)'' column is the coverage rate for the nominal 95\%
  confidence interval of the EstimNetDirected point and standard
  error estimates. Results are over 100 networks, each of which
  has 32 parallel estimation runs. $N_C$ is the number of networks for
  which a converged estimate was found (out of 100).  $\overline{N_R}$ is
  the mean number of runs that converged (out of 32).  Runs that did
  not converge are not included in the estimates.
  \end{flushleft}
  \label{tab:fnr_rate}
  \end{adjustwidth}
\end{table}

Table~\ref{tab:fpr_rate} shows the coverage and Type I error rates estimated from
simulated networks with zero effects. Note that in this case, coverage
and Type I error rate are effectively the same thing (as percentages, coverage
is $100 - \alpha$ where $\alpha$ is the Type I error rate) as the known
true value of the parameter is zero by design.
This table shows that the Type I error
rate in all but two cases is within the nominal 5\% range. In one
case, Matching Reciprocity for the categorical networks, the point
estimate of the Type I error rate is 9\% but the 95\% confidence
interval extends down to 5\%.  However the other case, Matching for
categorical networks, the point estimate of the Type I error rate is
11\% and the confidence interval extends down to only 6\%, so for this
particular case the Type I error rate is too high.  We note that the
coverage for this zero parameter is only 89\%, so it would appear that
our method is potentially subject to inferential error in such cases
where Matching Reciprocity is included in the model but the
corresponding baseline Matching parameter is not. We would recommend
not using such a model, and always including the corresponding
baseline parameters, as is standard practice in ERGM model building,
where configurations are nested within one another
\cite[Ch.~3]{lusher13}.

Also note that for two of the sets of simulated networks, the binary
node attribute networks with the Interaction or Reciprocity effects
set to zero, the number of runs and networks for which estimations
converged is very low (less than half). This is not problematic, as
these tests, where an effect not present in the network is included in
the model, could be considered instances of model mis-specification,
so the possibility of estimations not converging is to be expected.
Note that for estimations shown in Table~\ref{tab:fnr_rate}, where the
model is exactly correct (it is the same model that generated the
networks), converged estimates are obtained for all the simulated
networks.

\begin{table}
  \begin{adjustwidth}{-2.25in}{0in} 
    \centering
    \caption{{\bf Results from estimation of full network using
        EstimNetDirected estimating Type I error rate.}}
    \begin{tabular}{|r|l|l|r|r|r|r|r|r|r|r|}
\hline
N &  Attributes &  Effect &  Bias &  RMSE & estim. & lower  & upper & in C.I. (\%)    & $N_C$     & $\overline{N_R}$      \\
\thickhline
2000  &  Categorical  &   A2P-TD  & -0.0217 & 0.0657 &  1 &  0 &  5 & 99 & 100 & 31.94 \\ \hline
2000  &  Categorical  &   AinS  & -0.0017 & 0.0648 &  1 &  0 &  5 & 99 & 100 & 32.00 \\ \hline
2000  &  Categorical  &   AKT-T  & -0.0154 & 0.0837 &  0 &  0 &  4 & 100 & 100 & 32.00 \\ \hline
2000  &  Categorical  &   AoutS  & -0.0129 & 0.0706 &  1 &  0 &  5 & 99 & 100 & 32.00 \\ \hline
2000  &  Categorical  &   Matching  & 0.0239 & 0.0440 & 11 &  6 & 19 & 89 & 100 & 32.00  \\ \hline
2000  &  Categorical  &   MatchingReciprocity  & 0.1246 & 0.1981 &  9 &  5 & 16 & 91 & 100 & 32.00 \\ \hline
2000  &  Categorical  &   Reciprocity  & 0.4809 & 0.5493 &  2 &  1 &  7 & 98 & 100 & 30.86 \\ \hline
2000  &  Binary  &   A2P-TD  & -0.0143 & 0.0198 &  2 &  1 &  7 & 98 & 100 & 32.00 \\ \hline
2000  &  Binary  &   AinS  & -0.1234 & 0.1830 &  1 &  0 &  5 & 99 & 100 & 32.00 \\ \hline
2000  &  Binary  &   AKT-T  & -0.2473 & 0.5563 &  1 &  0 &  5 & 99 & 100 & 29.32 \\ \hline
2000  &  Binary  &   AoutS  & -0.0011 & 0.0954 &  0 &  0 &  4 & 100 & 100 & 32.00 \\ \hline
2000  &  Binary  &   Interaction  & -0.7966 & 3.0590 &  4 &  1 & 15 & 96 & 46 & 7.02 \\ \hline
2000  &  Binary  &   Receiver  & 0.0313 & 0.1577 &  5 &  2 & 11 & 95 & 100 & 31.33 \\ \hline
2000  &  Binary  &   Reciprocity  & -0.3127 & 1.2360 &  0 &  0 & 14 & 100 & 24 & 6.96 \\ \hline
2000  &  Binary  &   Sender  & 0.0244 & 0.1252 &  2 &  1 &  7 & 98 & 100 & 30.73 \\
\hline
    \end{tabular}
    \begin{flushleft}
  The ``estim.'', ``lower'', and ``upper'' columns show the point
  estimate and lower and upper 95\% confidence interval (C.I.),
  respectively, of the Type I error rate (false positive rate). This
  C.I. is computed as the Wilson score interval \cite{wilson27}.  The
  ``in C.I. (\%)'' column is the coverage rate for the nominal 95\%
  confidence interval of the EstimNetDirected point and standard error
  estimates. Results are over 100 networks, each of which has 32
  parallel estimation runs. $N_C$ is the number of networks for which
  a converged estimate was found (out of 100).  $\overline{N_R}$ is
  the mean number of runs that converged (out of 32).  Runs that did
  not converge are not included in the estimates.
  \end{flushleft}
    \label{tab:fpr_rate}
  \end{adjustwidth}
  \end{table}

\subsection*{Empirical network example}

Table~\ref{tab:pokec_results} shows a model estimated for the Pokec
online social network with the 20 highest degree hubs removed ($N =
1~632~783$).  This estimation took approximately 22 hours on cluster
nodes with Intel Xeon E5-2650 v3 2.30GHz processors using two
parallel tasks with 512 GB RAM each.

Regarding structural features, these results confirm centralization on
both in- and out-degree in the Pokec online social network. There is a
significantly positive reciprocity effect: friendships are more likely
to be reciprocated than not, conditional on the other features in the
model. There is also positive activity closure: people who send
friendship ties to the same people also tend to be friends. There is
also positive path (transitive) closure, in combination with negative
two-paths: friends of friends tend also to be friends. This can also
be interpreted in terms of the ``forbidden triad''
\cite{granovetter73}, in which an open two-path (of what we assume to
be ``strong'' ties as they represent friendship) is not closed
transitively. Such a triad is indeed less likely than by chance,
conditional on the other parameters in the model, in this network.

There is homophily on both region and age: people who live in the same
region are more likely to be friends than those who live in different
regions, and people of similar ages are more likely to be friends.
Interestingly, there is significant heterophily on the gender attribute:
people of different genders are more likely to be friends on this
online social network.

The convergence test plot for the Pokec online social network estimation
is shown in \nameref{S1_Fig}.

\begin{table}[!ht]
\centering
\caption{{\bf Parameter estimates for the Pokec online social network with hubs removed.}}
\begin{tabular}{|l|r|r|c|}
  \hline
Effect & Estimate & Std. error &  \\
\thickhline
Arc  & -20.642 & 0.381 &  *\\ \hline
Isolates  & -7.711 & 2.269 &  *\\ \hline
Reciprocity  & 33.473 & 3.834 &  *\\ \hline
Popularity spread (AinS)  & 1.720 & 0.128 &  *\\ \hline
Activity spread (AoutS)  & 1.900 & 0.280 &  *\\ \hline
Two-path (A2P-T)  & -0.014 & 0.003 & *\\ \hline
Shared popularity (A2P-D)  & 0.021 & 0.003 &  *\\ \hline
Shared activity (A2P-U)  & 0.022 & 0.003 &  *\\ \hline
Path closure (AKT-T)  & 2.151 & 0.411  & *\\ \hline
Popularity closure (AKT-D)  & 2.270 & 0.439  & *\\ \hline
Activity closure (AKT-U)  & 2.270 & 0.407 & *\\ \hline
Sender age  & 0.021 & 0.014 &  \\ \hline
Receiver age  & 0.022 & 0.013 &  \\ \hline
Diff age  & -0.099 & 0.012 &  *\\ \hline
Matching gender  & -1.164 & 0.287 &  *\\ \hline
Matching region  & 3.172 & 0.543 &  *\\
\hline
\end{tabular}
\begin{flushleft}
  Asterisks indicate statistical significance at $p < 0.05$.
\end{flushleft}
  \label{tab:pokec_results}
\end{table}

The algorithm parameters used for this estimation are shown in
\nameref{S1_Table} in the column for the Pokec network. As a general
guideline, we recommend using these parameters, with the exception of
\unix{EEsteps}. It may be useful to start with the default value of
1 000 (or even smaller) for this parameter, to relatively quickly obtain
initial results and check the trace plots for any obvious failure to
converge, such as a parameter value clearly diverging (due to a bad
model for instance). If there are no obvious problems then the
\unix{EEsteps} parameter can be increased if the t-ratios or trace
plots (as described in the section ``Convergence tests'') indicate
that the estimation has not yet converged.

\section*{Conclusion}

We have demonstrated an implementation of the EE algorithm for ERGM
parameter estimation capable of estimating models for social networks
with over one million nodes, which is far larger than previously
possible (without using network sampling). However there are several
limitations and scope for future work.
The implementation described here requires tuning some algorithm
parameters (\nameref{S1_Table}), however a simplified version of the
EE algorithm requires fewer parameters \cite{borisenko19} and may make
it easier to obtain converged models.

Although the use of hash tables to efficiently store the sparse
two-path matrices allows scalability to networks of millions of nodes,
it depends on sufficient sparsity of these matrices, and not all
empirical networks of interest satisfy this requirement. For example
the physician referral network described by An \etal \cite{an18a},
although having  approximately one million nodes, making it smaller than the
Pokec online social network, does not have a sufficiently sparse
two-path table for our implementation to work even with the largest
memory cluster node available to us (512 GB memory). Further work is
required to find a means of alleviating this problem.

Although the convergence heuristic plots we have described resemble a
goodness-of-fit test superficially, they are not actually
goodness-of-fit tests. The difficulty of generating (simulating)
numbers of very large networks from estimated ERGM parameters for
conventional goodness-of-fit tests for ERGM makes these methods
impractical for such large networks. One possibility to investigate is
to simulate snowball samples from the estimated parameters and compare
the distribution of network statistics of these simulated samples to
the corresponding distributions of network statistics in snowball
sample taken from the observed network. More generally, the idea of
such goodness-of-fit tests for extremely large networks may have to be
fundamentally re-examined, in light of the difficulty of any simple
model adequately fitting such a large network --- and the homogeneity
assumption that the same local processes operate across such a large
network may also no longer be realistic.

In addition, the ERGM MLE bias in canonical parameters may need to be
addressed by some bias correction technique, as was originally
explored for maximum pseudolikelihood estimation \cite{vanduijn09},
but does not appear to have been successfully pursued since for other,
now preferred (not least due to the results of \cite{vanduijn09}),
estimation methods.

An important next step is the strengthening of the theoretical basis
for the the EE algorithm. Existing commonly used methods for ERGM
parameter estimation are based on well-known algorithms with well
developed theoretical foundations such as stochastic approximation
\cite{snijders02} using the Robbins--Monro algorithm \cite{robbins51}
or MCMCMLE \cite{hunter06} based on the Geyer--Thompson method
\cite{geyer92} with increasingly sophisticated variants
\cite{hummel12}. In contrast, there are no theoretical guarantees
behind the EE algorithm \cite{byshkin18}, and although contrastive
divergence \cite{hinton02,asuncion10} and persistent contrastive
divergence \cite{younes88,tieleman08} are becoming widely used, more
theoretical work is required to understand their convergence
properties in general, and for ERGM parameter estimation in particular
\cite{fellows2014why}. Regarding the EE algorithm, the experiments
with simulated networks described in this paper give encouragement for
the validity and usefulness of this approach, but further work to
understand its convergence properties is a potentially fruitful area
of research \cite{borisenko19}.

\section*{Supporting information}


\paragraph*{S1 Appendix.}
\label{S1_Appendix}
 {\bf Algorithm descriptions.}
Pseudocode for the EE algorithm is
detailed below.  This algorithm uses an ERGM
sampler described immediately following it, the IFD sampler
\cite{byshkin16}, however other ERGM samplers may be used, and in
particular the ``basic'' sampler \cite{snijders02,wang09}, the
pseudocode for which is detailed in \cite[Supplementary Information]{byshkin18}
was used for the simulated network estimations.

In the algorithm descriptions, vectors such as $\boldtheta$,
$\boldsymbol{z}$, and $\boldsymbol{dz}$ have dimension equal to the
number of model parameters, $s$.  All vector operations are
elementwise, e.g. $\boldsymbol{dz}^2$ is the vector consisting of the
square of each element of $\boldsymbol{dz}$ and $\boldsymbol{D} \odot
\boldsymbol{dz}$ is the elementwise (Hadamard) product of $\boldsymbol{D}$ and
$\boldsymbol{dz}$ (a vector of the same dimension, $s$, as both
$\boldsymbol{D}$ and $\boldsymbol{dz}$).

  \begin{algorithmic}[1]
    \Require{$x_\mathrm{obs}$ is the observed graph, $\boldsymbol{\theta_0}$ is the initial parameter estimate, $\boldsymbol{D_0}$ is the initial derivative estimate. }
    \Ensure{Returned value $\boldsymbol{\theta_t}$ is the estimated parameter value.}
    \Statex
    \Function{EE}{$x_\mathrm{obs}, \boldsymbol{\theta_0}, \boldsymbol{D_0$}}
    \Let{$K_A$}{$10^{-4}$} \Comment{Multiplier of $\boldsymbol{D}$ to get step size multiplier}
    \Let{$c_1$}{$10^{-2}$} \Comment{Minimum magnitude of $\abs{\bar{\boldtheta}}$ (small positive constant)}
    \Let{$c_2$}{$10^{-4}$} \Comment{Multiplier of $\abs{\bar{\boldtheta}}/\sd{\boldtheta}$ to limit $\boldtheta$ variance}
    \Let{$K_{\mathrm{IFD}}$}{0.1} \Comment IFD sampler auxiliary parameter step  size multiplier
    \Let{$V$}{0} \Comment IFD sampler auxiliary parameter
    \Let{$M_{\mathrm{outer}}$}{1000} \Comment{Steps of Algorithm EE}
    \Let{$M_{\mathrm{inner}}$}{100} \Comment{Inner iterations of Algorithm EE}
    \Let{$m$}{1000} \Comment{Number of sampler iterations}
    \Let{$t$}{0}
    \Let{$x$}{$x_\mathrm{obs}$}
    \Let{$\boldsymbol{D}$}{$\boldsymbol{D_0}$}
    \Let{$\boldsymbol{dz}$}{$\boldsymbol{0}$} \Comment{Vector of accumulated change statistics}
    \For{$i \gets 1 \mathrm{~to~} M_{\mathrm{outer}}$}
      \For{$j \gets 1 \mathrm{~to~} M_{\mathrm{inner}}$}

      \Let{$(\boldsymbol{dzAdd}, \boldsymbol{dzDel})$}{\Call{Sampler}{$x, \boldtheta_{\boldsymbol{t}}, m, K_{\mathrm{IFD}}, V$}}
      \Let{$\boldsymbol{dz}$}{$\boldsymbol{dz} + \boldsymbol{dzAdd} - \boldsymbol{dzDel}$} \Comment{Accumulate accepted change statistics}      
        \Let{$\boldtheta_{\boldsymbol{t+1}}$}{$\boldsymbol{\theta_t} - \sign{\boldsymbol{dz}} \odot K_A  \boldsymbol{D} \odot {\boldsymbol{dz}}^2$}
        \Let{$t$}{$t + 1$}
     \EndFor
     \Let{$\boldsymbol{D}$}
         {$\boldsymbol{D} \odot \left[ c_2
             \frac{\max \left(
               \abs{\overline{\boldtheta_{\boldsymbol{t-M_{\textbf{inner}} \leq k < t}}}},
               c_1 \right) }
                  {\sd{\boldtheta_{\boldsymbol{t-M_{\textbf{inner}} \leq k < t}}}}
                  \right]^{\frac{1}{2}}$}
     \EndFor
     \State \Return{$\boldsymbol{\theta_t}$}
    \EndFunction
  \end{algorithmic}

The algorithm to sample from ERGM distributions with
Metropolis--Hastings using the IFD sampler \cite{byshkin16} is
described below. Note that the Arc parameter $\theta_L$ must not be
included in the model when using the IFD sampler, instead it is
calculated from the IFD sampler auxiliary parameter $V$ as
\begin{equation}
  \theta_L = V - \log \left( \frac{L_{\mathrm{max}} - L_{\mathrm{obs}}}{L_{\mathrm{obs}}+1} \right)
\end{equation}
where $L_{\mathrm{max}} = N(N-1)$ is the number of possible arcs in a directed
graph and $L_{\mathrm{obs}}$ is the number of arcs in the observed graph
$x_{\mathrm{obs}}$.

  \begin{algorithmic}[1]
    \Require{$x$ is a directed graph, $\boldtheta$ is vector of parameters, $m$ is number of sampler iterations, $K_{\mathrm{IFD}}$ is the multiplier for the auxiliary parameter step size, \eg $K_{\mathrm{IFD}} =0.1$. Initially set IFD auxiliary parameter $V = 0$.}
    \Ensure{Return value ($\boldsymbol{dzAdd}, \boldsymbol{dzDel}$) accumulated change statistics of accepted (add, delete) moves. The graph $x$ is updated by the accepted moves and the IFD auxiliary parameter $V$ is updated.}
    \Function{Sampler}{$x, \boldtheta, m, K_{\mathrm{IFD}}, V$}
    \Let{isDelete}{False}
    \Let{$N_\mathrm{add}$}{0} \Comment number of add moves
    \Let{$N_\mathrm{del}$}{0} \Comment number of delete moves
    \Let{$\boldsymbol{dzAdd}$}{$\boldsymbol{0}$}
    \Let{$\boldsymbol{dzDel}$}{$\boldsymbol{0}$}
    \For{$w \gets 1 \mathrm{~to~} m$}
    \If{isDelete} \Comment Delete move
        \Let{$N_{\mathrm{del}}$}{$N_{\mathrm{del}} + 1$}      
        \State Choose two nodes $i,j$ ($i \neq j$) with arc $i \rightarrow j$ uniformly at random    
    \Else \Comment Add move
        \Let{$N_{\mathrm{add}}$}{$N_{\mathrm{add}} + 1$}                  
        \State Choose two nodes $i,j$ ($i \neq j$) with no arc $i \rightarrow j$ uniformly at random
    \EndIf
      \State Compute change statistic $dz_A$ for adding arc $i \rightarrow j$ (if $\lnot$ isDelete) or deleting arc $i \rightarrow j$ (if isDelete) for each statistic $A$
      \State $V_s \gets -1$ if isDelete else 1
      \Let{$\alpha$}{$\min \left\{ 1,  \exp \left( \sum_A \left[ \theta_A  dz_A \right]  + V_s  V \right)\right\}$} \Comment proposal acceptance probability
      \If {$\mathrm{Unif}(0,1) < \alpha$} \Comment{Accept change with probability $\alpha$}
        \If{isDelete}
          \Let{$\boldsymbol{dzDel}$}{$\boldsymbol{dzDel} - \boldsymbol{dz}$}
          \Let{$x_{ij}$}{0}        
        \Else
          \Let{$\boldsymbol{dzAdd}$}{$\boldsymbol{dzAdd} + \boldsymbol{dz}$}
          \Let{$x_{ij}$}{1}    
          \EndIf
        \Let{isDelete}{$\lnot$ isDelete}
      \EndIf
      \EndFor
      \Let{$V_{\mathrm{step}}$}{$\frac{(N_{\mathrm{del}} -  N_{\mathrm{add}})^2}{(N_{\mathrm{del}} +  N_{\mathrm{add}})^2}$}
      \If{$N_{\mathrm{del}} -  N_{\mathrm{add}} > 0$}
        \Let{$V$}{$V - K_{\mathrm{IFD}}  V_{\mathrm{step}}$}
      \Else
        \Let{$V$}{$V +  K_{\mathrm{IFD}}  V_{\mathrm{step}}$}        
      \EndIf
      \If{$\frac{\abs{N_{\mathrm{del}} -  N_{\mathrm{add}}}}{N_{\mathrm{del}} +  N_{\mathrm{add}}} > 0.8$}
        \State Warn that $K_\mathrm{IFD}$ might be too small.
      \EndIf
    \State \Return{$(\boldsymbol{dzAdd}, \boldsymbol{dzDel})$} 
    \EndFunction
  \end{algorithmic}

\paragraph*{S1 Table.}
\label{S1_Table}
 {\bf EstimNetDirected parameter settings.}
\begin{adjustwidth}{-2.25in}{0in} 
  \centering
\begin{tabular}{|l|l|l|r|r|}
\hline
Parameter name & Pseudocode        & Simulated & Pokec  \\
\thickhline
ACA\_S         & $K1_A$             &  0.1        & 0.1        \\ \hline
ACA\_EE        & $K_A$              &  $10^{-9}$   & $10^{-7}$   \\ \hline
compC          & $c_2$              &  0.01       & 0.01       \\ \hline
samplerSteps   & $m$                & 1000        & 1000       \\ \hline
Ssteps         & $M1$               & 50          & 1000       \\ \hline
EEsteps        & $M_{\mathrm{outer}}$  & 500          & 1500       \\ \hline
EinnerSteps    & $M_{\mathrm{inner}} $ & 100          & 100        \\ \hline
useIFDsampler  &   ---             & False        & True       \\ \hline
ifd\_K         & $K_{\mathrm{IFD}}$   & ---          & 0.1        \\ \hline
\end{tabular}
\begin{flushleft}
``Pseudocode'' is the notation used for the parameter in
  \nameref{S1_Appendix} or \cite[Supplementary Information]{byshkin18}.
  ``Simulated'' is the value used for estimating the simulated networks
  and ``Pokec'' is the value used in estimating the model for the
  Pokec online social network.
\end{flushleft}
\end{adjustwidth}

\paragraph*{S1 Fig.}
\label{S1_Fig}
      {\bf Convergence test plot for Pokec (hubs removed) estimation.} The observed
      network statistics are plotted in red with the statistics of the
      EE algorithm simulated networks on the same plot as black
      boxplots, or blue on histogram plots. Note that on the triad
      census plots, triads 003, 012, and 102 are omitted as the
      extremely large counts cause numeric overflow in the
      \unix{igraph} library \cite{csardi06} for a network this large.
      
\includegraphics[width=\textwidth]{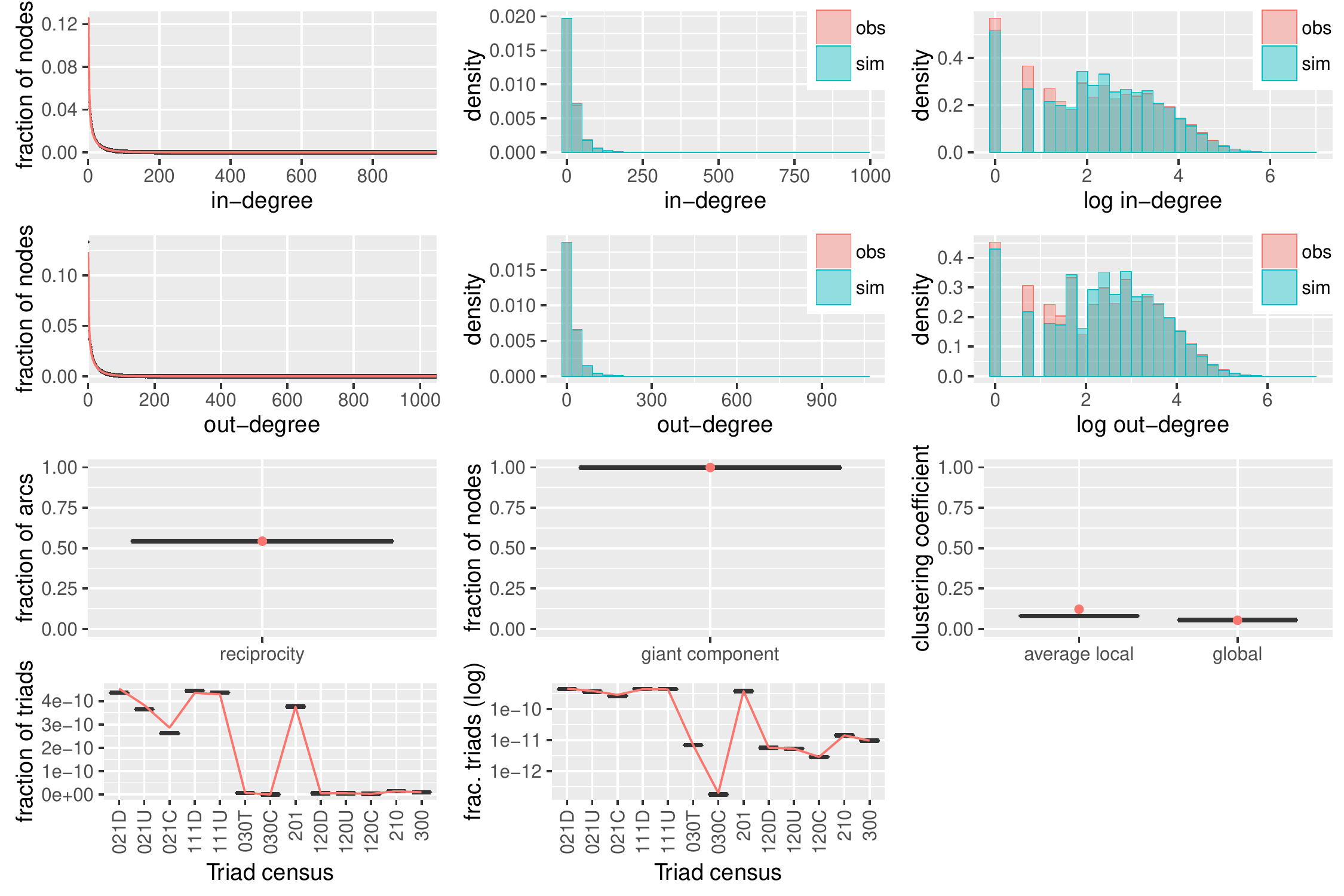} 

\section*{Acknowledgments}

We gratefully acknowledge the support of Swiss National Science
Foundation NRP 75 Big Data project 167326 ``The Global Structure of
Knowledge Networks: Data, Models and Empirical Results''.  We thank Dr
Pavel Krivitsky, Prof.~Dean Lusher, Prof.~Antonietta Mira, and Prof.~Ernst Wit for
useful discussions.

\nolinenumbers

%
%
%


\end{document}